# ENHANCED ANTENNA POSITION IMPLEMENTATION OVER VEHICULAR-AD HOC NETWORK (VNET) IN 3D SPACE

## Soumen Kanrar[1] and Mohammad Siraj[2]

[1]Department of Computer Engineering ,College of Computer and Information Sciences, King Saud University, Riyadh -11543, SA

soumen_kanrar@yahoo.co.in

[2]Department of Computer Engineering ,College of Computer and Information Sciences, King Saud University, Riyadh -11543, SA

siraj@ksu.edu.sa

## ABSTRACT

*The technology related to networking moves wired connection to wireless connection. The basic problem concern in the wireless domain, random packet loss for the end to end connection. In this paper we show the performance and the impact of the packet loss and delay, by the bit error rate throughput etc with respect to the real world scenario vehicular ad hoc network in 3-dimension space (VANET in 3D). Over the years software development has responded to the increasing growth of wireless connectivity in developing network enabled software. In this paper we consider the real world physical problem in three dimensional wireless domain and map the problem to analytical problem .*
*In this paper we simulate that analytic problem with respect to real world scenario by using enhanced antenna position system (EAPS) mounted over the mobile node in 3D space. In this paper we convert the real world problem into lab oriented problem by using the EAPS –system and shown the performance in wireless domain in 3 dimensional space.*

## KEYWORDS

*EAPS,    VANET,    Bit error rate,    throughput,    antenna,    wireless.*

## 1 Introduction

Vehicular ad hoc networks (VANET) differ from usual mobile ad hoc networks (MANETs) in many different aspects. First VANET consist of highly mobile nodes which moves in the same or opposite directions. The node moves along the two different paths or in a curve . In a shared wireless medium, during the communication establishment between the mobile nodes, blindly broadcasting packets may lead to frequent contention and congestion among neighboring nodes. This problem is some time referred to as the broadcast storm problem while multiple solution exit to alleviate the broadcast storm in the usual MANET environment. In this paper we consider particular types of VANET problem in 3 dimensional spaces. Here we consider the vehicle as the mobile node. The vehicle receives alert message from the ground station and this message is to be relayed to the neighboring vehicles. The received packets will be relayed by the vehicle which is in high speed motion in 3 dimension space. The distance based schemes used in vehicular ad hoc network are weighted p-persistence Broadcasting, slotted 1-persistence broadcasting and slotted p-persistence to find a traffic flow pattern for the vehicular ad hoc network [1]. The performance for the mobile node in ad hoc network by broadcasting the packets scales at the receiver node [2].Various threshold-based techniques have been implemented to solve the broadcast storm problem in mobile domain [3] such as counter–based, distance based and location based. Directional antenna is used to mitigate broadcast redundancy and alleviate contention at the MAC layer [4].Various mathematical models are used to reduce the redundancy in the broadcast storm problem for mobile Ad hoc network [5]. Real world flight data is used in 3 dimensional spaces to validate by simulating vehicular ad hoc network [6]. By mounting isotropic antenna over the mobile node, performance of the network is achieved by message exchange in context client – server connection in wireless domain [7]. In [8] authors have considered Realistic "Energy consumption" model in physical layer for a device in mobile ad hoc network .In our work the proposed scheme don't require a node to keep track of its neighbors.

## 2. Proposed Model

In our proposed real time problem we consider "free space propagation model" . The aircraft is a receiver $R_x$ (consider as mobile node). It receives message from the ground station in the 3D space via the figure -1 and figure-1.1 . The receiver (ie aircraft) is in an arbitrary height and receives the message from the ground station $T_x$ (fixed base station as the transmitter). We mount three antennas (directional, isotropic and cone antenna) over the base station i.e ground station and isotropic antenna over the aircraft. Here the aircraft moves in a particular trajectory from the earth surface. There should be a minimum distance between every aircraft. There are more than one aircraft at a particular height , maintaining a minimum distance among them. If we consider the aircraft as the vehicle and the ground station as the fixed base station, then clearly it is vehicular ad hoc network (VANET). Now the base station broadcast the alert massage and that alert massage is to be relayed between the neighbor aircrafts. Due to broadcasting from the fixed base station and from the relayed packet by neighbor's aircraft ( using the fixed channel frequency) , the problem of packet collection and contention for random broadcasting arises. Redundant broadcast is also an issue for this problem scenario. In this paper we show the system performance by measuring bit error rate and throughput by using "Enhanced antenna position system (EAPS)" for various types of antennas mounted over the mobile node.

Diagram for the proposed Model

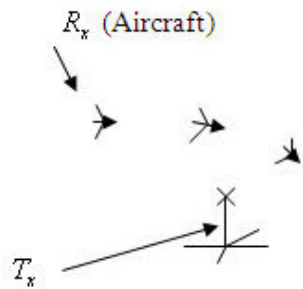

Figure 1

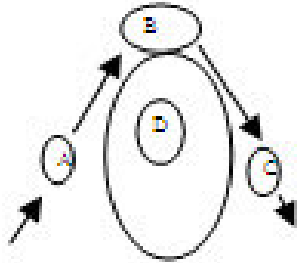

Figure 1.1

Figure-1 represent the movement of the aircraft's and figure-1.1 represent the different position of the vehicle in its motion. According to figure 1.1 points A, B, C represent the position of the vehicle and D is the transmitter (Ground station)

## 3 Preliminaries

*Manet*: (Mobile ad hoc network) consist of a set of mobile hosts that may communicate with one another from time to time. No base station is supported. Each host is equipped with a CSMA/CA (carrier sense multiple access with collision avoidance) transceiver .In such a environment a host may communicate with another directly or indirectly.

*Redundant rebroadcast*: When a mobile host decides to rebroadcast message to its neighbors which already have the message.

*Contention*: After a mobile host broadcast a message, if many of its neighbors decide to rebroadcast the message, these transmission ( which are all from nearby hosts) may severely cause contention with each other.

*Collision:* Due to deficiency of back off mechanism in mobile domain , the lack of RTS/CTS dialogue and the absence (CD) collision detection, collision is more likely to occur and cause more damage in vehicular ad hoc network.

*Directional Antenna:* A directional antenna such as a parabolic antenna attempt to radiate most of its power in the direction of a known receiver.

*Isotropic antenna*: Isotropic antenna is an antenna that transmits equally in all directions.

*Cone antenna*: A conventional discone antenna is a cone antenna. This antenna typically has a feed structure that is within the cone. The cone antenna is used in wireless local area network (WLAN).

*Antenna gain*: It is power output, in a particular direction, compared to that produced in any direction by a perfect omni directional antenna (isotropic antenna).Antenna gain is a measure in dB that how much more power an antenna will radiate in a certain direction with respect to that which would be radiated by a reference antenna. The flowing equation represents the relationship between antenna gain and effective area,

$$G = \frac{4\pi A_e}{\lambda^2} = \frac{4\pi f^2 A_e}{c^2}$$

Where, G = Antenna gain, $A_e$ = Effective area, f = Carrier frequency, c = speed of light

$\lambda$ = Carrier wavelength.

*Azimuth angle*: The Azimuth angle, often denoted with a $\theta$ is the angle that the direct transmission makes with respect to a given reference angle (often the angle of the target receiver) when looking down on the antenna from above.

*Elevation angel*: The elevation angle is the angle that the transmission direction makes with the ground elevation angle and is denoted with a $\phi$.

*Free space model*: The free space propagation model assumes the ideal propagation condition that there is only one line of –sight path between the transmitter and receiver.

H.T.Friis presented the following equation to calculate the received signal power in the free space at distance d from the transmitter. The analytic form of free space model is

$$P_r(d) = \frac{P_t G_t G_r \lambda^2}{(4\pi)^2 d^2 L}$$

$P_t$ is the transmitted signal power, $G_t$ and $G_r$ are the antenna gains of the transmitter and receiver respectively, L $(1 \leq$ L$)$ is system loss and $\lambda$ is the wavelength.

*Roll*: Rotation around the front –to –back axis is called roll via fig- 2

*Pitch*: Rotation around the side –to – side axis is called pitch via fig- 2

*Yaw*: Rotation around the vertical axis is called yaw via fig -2

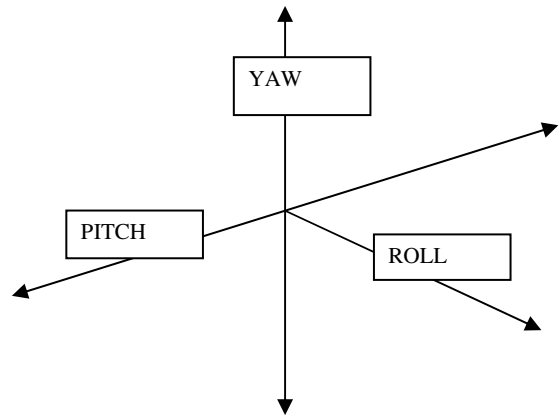

Figure – 2

## 4 System Model:

The "Enhanced Antenna positioning" model, includes all six variables need to describe the mobile node in 3-D space." latitude, longitude, altitude, yaw( bearing) direction of traveling pitch(angle of ascent) and roll ( rotation angle) at the node position value attributes. In addition, the model incorporates Antenna position values (Theta) $\theta$ , horizontal angle (phi) $\phi$ and rotation angle describing the attitude of the antenna mounted on the node. The model also includes modified receive and transmit antenna gain pipeline stages that utilize these new variables to describe the motion of an antenna mounted on a mobile node.

The altitude information of the mobile node is often reported in terms of roll, pitch and yaw. Facing is a combination of the nodes yaw and geospatial facing.

Facing calculated by latitude and longitude and yaw value from the attitude.

The formula to determine node's facing is

F= arctan( $X_{n+1} - X_n, Y_{n+1} - Y_n$ ) + $yaw_n$ .

Here $X_{n+1}$ is the next longitude position.

$X_n$ is the current longitude.

$Y_{n+1}$ is the next latitude position.

$Y_n$ is the current latitude position.

$yaw_n$ is the mobile node rotation around its z axis.

The pointing direction of the antenna is assumed to coincide with the positive z-axis of the antenna gain pattern.

For calculating pointing direction s the antenna to the mobile node (receiver which is vehicle or aircraft via fig-1) is fixed. As the aircraft changes altitude and position, so does the pointing direction of the antenna. The vector that identifies this pointing position defines the azimuth and elevation angles that map to an antenna gain in the antenna pattern. In 2 dimensional space we consider altitude remain fixed but the position is changing.

The antenna gain pipeline stages were modified to make use of the node's attitude information. First the antenna mounted and pointing direction attributes are read from the antenna module. Each value, antenna mount and pointing direction is stored as three angles, phi theta and rotation ($\phi, \theta \psi$). The antenna mount values are used to create two vectors one for the direction of the antenna mounted on the platform and the other for the "up" value of the antenna's bore sight. The two vectors are modified by the node's facing, pitch and roll. The facing pitch and roll are described as three angles each representing the degree of rotation around the Z-axis, Y-axis and node's axis. Once the antenna vectors have been modified, the antenna actual pointing direction in the simulation's in 3D space is known.

The antennas pointing direction is used to determine the antenna gain from the antenna pattern.

## 5 Simulation Model

OPNET was used to build the simulation model. All the operations are done by using OPNET kernel procedures. This is Baseline simulation .The Antenna gain pipeline stage is modified at the transmitter and receiver according to the model requirement. The role of the pipeline stage is to compute the antenna gain, throughput and Bit error rate, for the different position of the mobile node.

The Enhanced Antenna position system (EAPS) is used to improve the performance gain. In general EAPS uses two modes

1) Fixed to object
2) Locked to object

*Fixed to object*: The fixed to object mode is describing the movement of an antenna that mounted over the mobile node and moves with the mobile node.

*Locked to Target*: In this mode the antenna always points at a specific target.

In both the modes, rotation angle parameters are used so the antenna rotates about its pointing axis *ref-* (10).

### 5.1 Scenario Description

The system consists of a fixed transmitter and mobile node receiver with mobile Jammer moving in a trajectory in the area around 8000x4000 meters. The receiver is model by using random waypoint mobility i.e. the receiver moves randomly. The jammer node is used to create radio noise. Here the transmitter and receiver modules use different channel for connection setup by using the control packet . For message broadcasting and message forwarding the transmitter and the receiver module use the data packet. The jammer is continuously changing its position around the receiver by moving in a trajectory. The simulation is allowed to run for 12 minutes .

The size of the packet send by the transmitter is 1024 bits. The transmitter broadcast each packet in 1 minute. The transmission power at the transmitter is 20watts.The receiver node moves with velocity of 10meter/second.

## 5.2 Node model

**5.2.1 Transmitter :** The transmitter composed of three modules via figure -3
1) Simple source i.e. packet generator.
**2)** Antenna target tracker module. This module tracks the antenna location.
**3)** Radio transmitter module. This module transmits the packets on a radio channel.

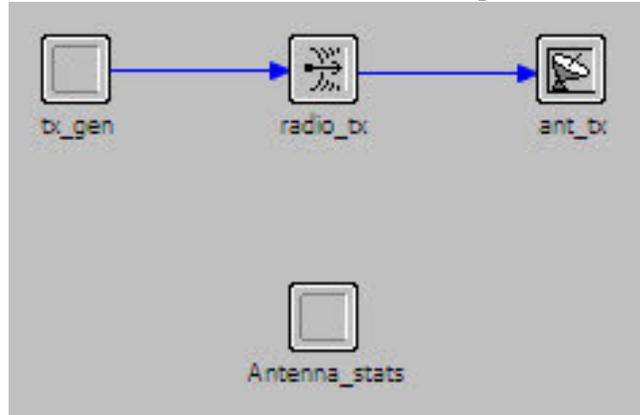

Figure -3

**5.2.2 Receiver:** The receiver composed of three modules via figure- 4
1) Antenna module.
2) Radio receiver module: This module defined the gain which will be adjusted at run time.
3) Sink processor module: This module store the received packets.

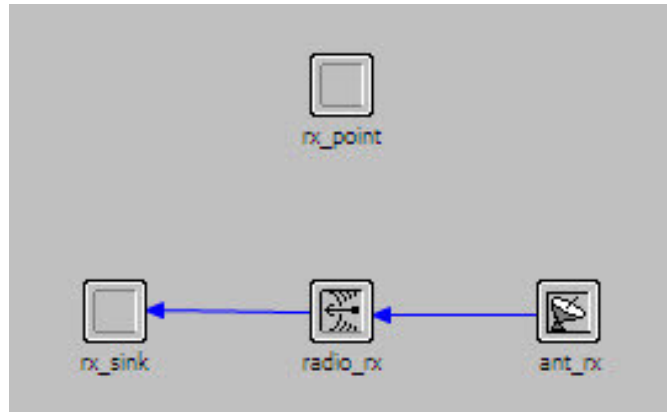

Figure -4

### 5.3 Summarizes the simulation parameters

Table-1

| Transmission power | 20 w |
|---|---|
| Simulation time | 12m |
| Movement area | 8x4 km |
| Node Movement speed | 10 meter/sec |
| Modulation | BPSK |
| Packet size | 1024 bits |
| Packet interval  Time (broadcast interval) | 1.0m |

## 6 Performance analyses:

For the performance of the system, in this paper we study the bit error rate and the throughput for the different types of antenna used. Isotropic, directional antenna, full cone antenna were mounted over the transmitter and the isotropic antenna over the receiver mobile node (vehicle). We move  the jammer in a trajectory that artificially improves the situation as real world.The jammer and the transmitter produce packet independently and asynchronously.

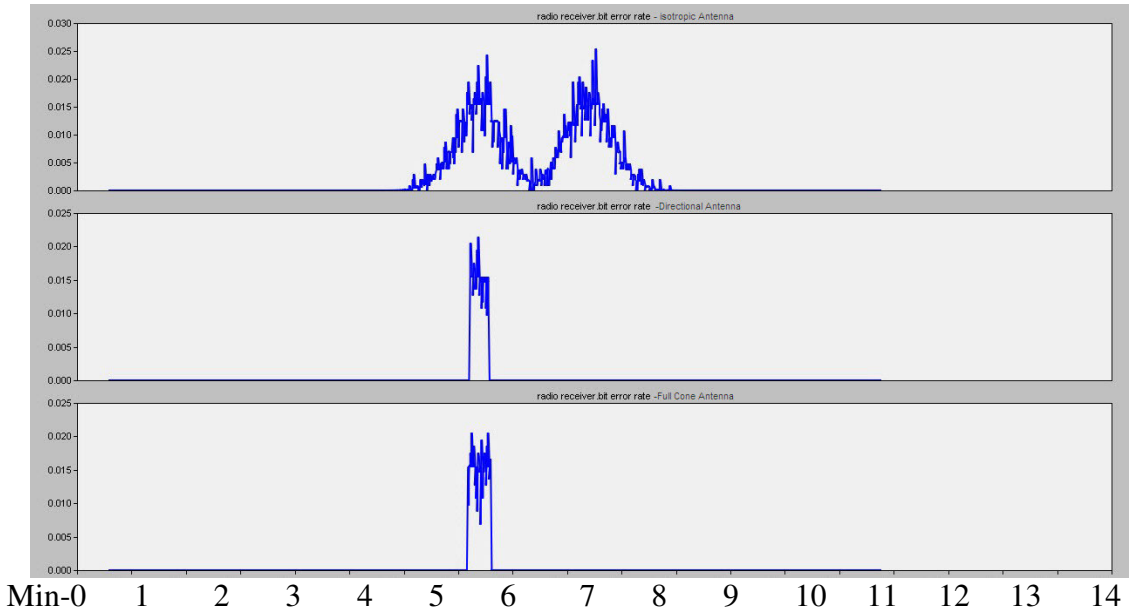

Figure-5  Bit error rate with no antenna tracker

Bit error rate is concerned  about the rejection of packet, which the vehicle receives with error. Whereas  in the case of throughput, how many packets are  received properly with respect to the packet transmitted from the transmitter in presence of the jammer. The jammer artificially generate noise packet according to the figure- 5 and figure -6. The simulation is done by packet to packet basis. In the case of received power, we compare  signal strength at the receiver side to the transmitted signal power. The jammer artificially generate noise as in  real life scenario.

Figure -5 and figure -6 represent the bit error rate without antenna tracker and with antenna tracker in context of figure 1.1

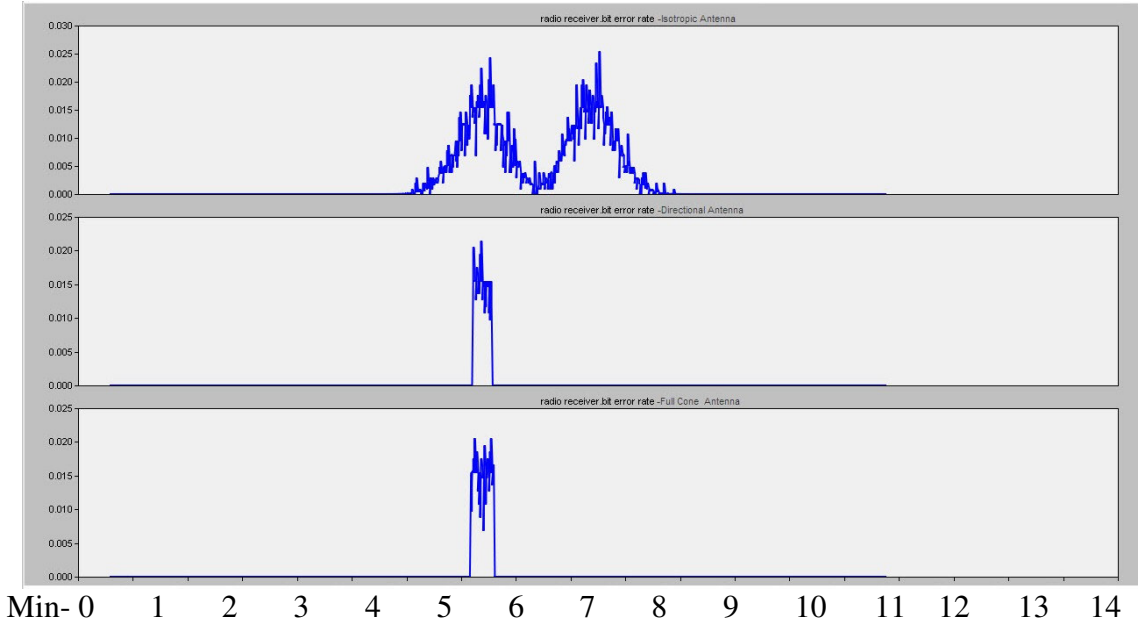

Min- 0   1   2   3   4   5   6   7   8   9   10   11   12   13   14

Figure -6  Bit error rate with antenna tracker

Bit error rate initially is zero at the receiver node (vehicle) because  the distance between the jammer node and receiver node is large.  After 5.5 minutes the direction vector between the jammer antenna and the receiver antenna is in  line with the direction of gain of  receiver antenna. The receiver node (vehicle) get maximum bit error during the interval 6.5 to 8.5 minutes and 9.0 to 11.0 minutes for the isotropic  antenna via –figure 5 and figure-6 and the interval 6.5 to 7.5 for other types of antenna which are mounted over the receiver node (vehicle) via figure -5 and figure-6.

Figure -7 and figure -8 represent the throughput of the system with antenna tracker and without antenna tracker in context of figure-1.1. Clearly it seen from the figure during the time interval 6.5 to 8.5 minutes and during the interval 9.0 to 11.0 minutes the throughput  approaches to zero for the isotropic antenna mounted over the receiver vehicle, where the Bit error rate is maximum. During the interval 6.5 to 7.5 minutes throughput approaches to zero for the other types of antenna mounted over the receiver mobile vehicle.

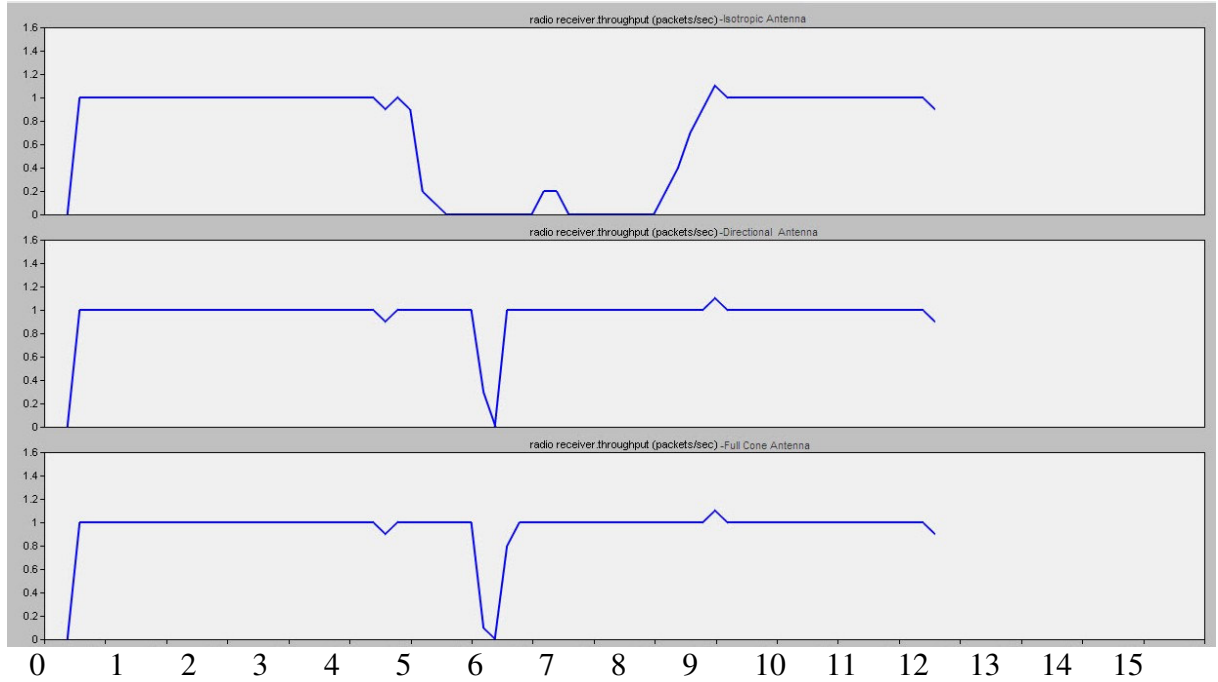

Figure-7 Throughput with no antenna tracker

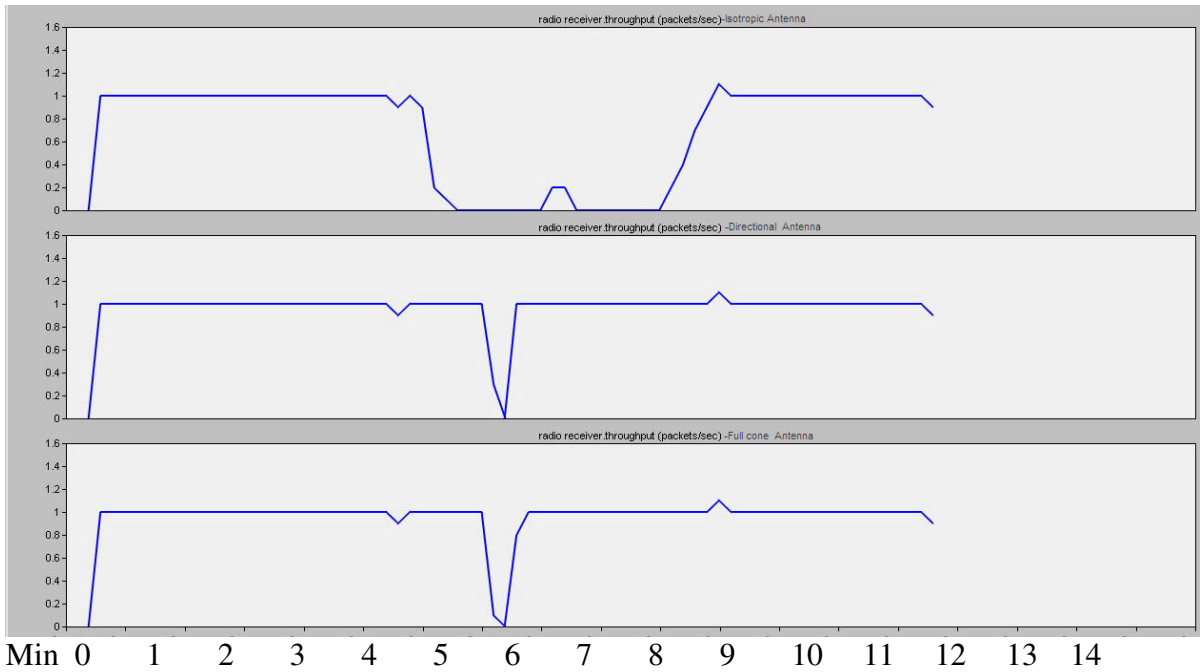

Figure-8 Throughput with antenna tracker

In the figure -7 and figure -8, initially the throughput is maximum for all types of antenna mounted over the receiver mobile vehicles but it decreases to zero with the increase of time (at 6.5 minute). Throughput almost approaches to maximum 1.0 and the Bit error rate is maximum via figure- 5 and figure- -6 (0.025 errors/ bits).

## 7.Concluding Remarks:

In this paper we have considered the real life problem in vehicular ad hoc networks in 3 dimension space. In this paper we have shown the performance of the vehicular ad hoc network in 3 dimensional space with bit error rate and throughput using OPNET simulator. This paper can be further extended to measure the broadcast redundancy and the zone wise message broadcasting with the help of clustering in high speed ad hoc networking.

For further work, to get better performance evaluation there is a need to create complex simulation scenarios. Algorithms have to be developed and the complex traffic pattern taken into consideration.

## REFERENCES:

[1]     Broadcast storm Mitigation Techniques in Vehicular Ad hoc Networks,N.Wisitpon, Gphan and .K.Tonguz,J.S.Parikh,P.Mudalige, F.Bai,V.Sadekar IEEE Wireless communications December - 2007 (Open call)

[2]     Broadcast Reception rates and effects of priority access in 802.11 based Vehicular-Adhoc networksM.Torrent –Moreno, D.Jiang and H.Hartenstien Proc.ACM Int'l wksp Vehicular Ad hoc Networks – Philadelphia PA Oct 2004

[3].    S.N. et al " The Broadcast storm problem in a mobile Ad hoc Network " Proc . ACM int'l conf .Mobile computing and Networking – SEATTLE WA-1999.

[4]     C. Hu, Y. Hong and J. Hou " On Mitigating the Broadcast storm problem with directional antennas" Proc IEEE ICC vol-1 Seattle WA May 2003

[5]     SZE –Yao Ni , Yu –Chee Tseng ,Yuh-Shyan,Jang- ping Sheu "The Broadcast storm problem in a mobile Ad hoc Network" ACM -1999.

[6]     R.Preston, J Doane, D.Kiwior "Using real world data to create OPNET Models DRAFT" The - MITRE Corporation

[7]     Optimizing the performance of MANET with Enhanced Antenna position system Jackline Alphones and Mohamed Naufal M.Saad IJCSNS -2009 Vol -9 No -4.

[8]     ."Simulating mobile ad hoc networks in city Scenarios" IIya Stepanov, Kurt Rothermel Elsevier-Computer, communication -2007.

[9].    "Transmission Scheduling in Ad hoc Networks With Directional Antennas "L.Bao and J.J Garcia – Luna- Aceves ACM/Sigmobile Mobicom,28sep 2002.

[10]    Arne Schmitz, Martin Wenig "The Effect of the Radio Wave Propagation Model in mobile Wireless Networks" MSWiM 2006 ACM

[11]    Illya Stepanov, Kurt Rothermel ,On the impact of a more realistic physical layer on MANET simulation results ScienceDirect Wireless Networks 6 (2008) pp 61-78

[12]    Gianni A. Di caro, Fredrick Ducatelle, and Luca m. Gambardella "A Simulation study of routing performance in realistic urban scenarios for MANETS "– Proceedings of ANTS - 2008

[13]    Models for Realistic Mobility and radio wave propagation for Wireless Networks Simulation By Mesut Gunes and Martin Wenig Chapter-11 "Guide to Wireless Wireless Network" Springer - 2009

[14]    Ratish J. Punnose,Pavel V. Nikitin, and Daniel D. Stancil " Efficient Simulation of Rician Fading
        within a Packet Simulator". IEEE Vehicular Technology Conference, Sept 2000.

[15]    www.opnet.com

**Authors**

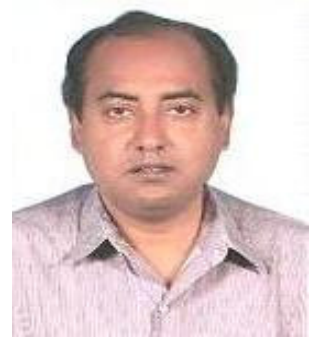

**Soumen Kanrar** received the M.Tech. degree in computer science from Indian Institute of Technology Kharagpur India in 2000. Advanced Computer Programming RCC Calcutta India 1998.  and MS degree in Applied Mathematics from Jadavpur University India in  1996.
During 2001-2003 he worked in Techno India as faculty in computer-engineering department. During 2003-2005- he worked in Dr. B.C.Roy Engineering college India as faculty in the Computer Science and Engineering department. During 2005-2008 he worked  in Durgapur Institute of Advanced Technology India  as faculty in the Department of Computer Science and Engineering. Currently working as a Research Associate in the College of Computer & Information Sciences, King Saud University, Riyadh, Saudi Arabia

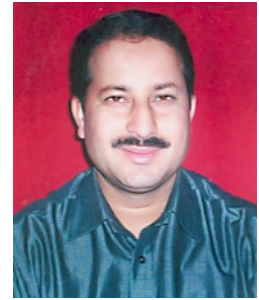

**Mohammad Siraj** received his M.E degree in Computer Technology and Applications from Delhi College of Engineering ,Delhi-India in 1997 and BE degree in Electronics and Communication Engg from Jamia Millia Islamia, New Delhi in 1995. He has worked as Scientist in Defence Research and Development Organisation (DRDO) Ministry Of Defence, India from 1995- 2000.  Currently he is working as a Lecturer as a  Lecturer in the College of Computer & Information Sciences, King Saud University, Riyadh, SA